\documentclass[%
 reprint,
 showpacs,
 amsmath,amssymb,
 aps,
]{revtex4-1}
\usepackage{amsmath}
\usepackage{cases}
\usepackage{amssymb}
\usepackage{CJK}% Chinese character support
\usepackage{graphicx}% Include figure files
\usepackage{dcolumn}% Align table columns on decimal point
\usepackage{bm}% bold math
\usepackage{color}
\usepackage[pagewise]{lineno}% add line numbers

\begin{document}
\begin{CJK*}{GB}{gbsn} % Use default fonts from CJK (see below)

\preprint{APS/123-QED}

\title{Experimental Determination of One- and Two-Neutron Separation Energies \\
for Neutron-Rich Copper Isotopes
}% Force line breaks with \\
\author{Mian YU(ÓÚÃã)$^{1,2}$}
\author{Hui-Ling WEI (κ»ÛÁá)$^{1}$}
\author{Yi-Dan SONG (ËÎÒ»µ¤)$^{1}$}
\author{Chun-Wang MA(Âí´ºÍú)$^{1,3}$}
\thanks{Email address: machunwang@126.com}

%\homepage{http://www.researcherid.com/rid/B-3848-2009}
\affiliation{
$^{1}$ Institute of Particle and Nuclear Physics, Henan Normal University, \textit{Xinxiang 453007}, China\\
$^{2}$ School of Biomedical Engineering, Xinxiang Medical University, \textit{Xinxiang 453003}, China\\
$^{3}$ Shanghai Institute of Applied Physics, Chinese Academy of Sciences, \textit{Shanghai 201800}, China\\
}


%\affiliation{
%\affiliation{

\date{\today}
%Start preparation 6th, Oct. 2016., First draft 9th, Oct. 2016.

%\linenumbers% show line numbers
\begin{abstract}
A method is proposed to determine the one-neutron $S_n$ or two-neutron $S_{2n}$ separation energy  of neutron-rich isotopes. Relationships between $S_n$ ($S_{2n}$) and isotopic cross sections have been deduced from an empirical formula, i.e., the cross section of an isotope exponentially depends on the average binding energy per nucleon $B/A$. The proposed relationships have been verified using the neutron-rich copper isotopes measured in the 64$A$ MeV $^{86}$Kr + $^{9}$Be reaction. $S_n$, $S_{2n}$, and $B/A$ for the very neutron-rich $^{77, 78, 79}$Cu isotopes are determined from the proposed correlations. It is also proposed that the correlations between $S_n$, $S_{2n}$ and isotopic cross sections can be used to find the location of neutron drip line isotopes.
\end{abstract}

                             % 25.70.Pq Multifragment emission and correlations
                             % 24.60.-k Statistical theory and fluctuations
                             % 25.70.Mn Projectile and target fragmentation
                             % 21.65.Cd Asymmetric matter, neutron matter
%Use showkeys class option if keyword
\maketitle
\end{CJK*}

%\tableofcontents
%\linenumbers% show line numbers

\section{Introduction}
The rare isotopes, including the very neutron-rich isotopes and the very proton-rich ones, which are near the neutron and proton drip lines, consistently attract the interest of both theoretical and experimental scientists. Indicated by the easy separation of neutrons, isotopes near the neutron drip line have very small one-neutron separation energy ($S_n$) or two-neutron separation energy ($S_{2n}$), which means that the one or two neutrons can be removed from the nucleus quite smoothly. The same happens in the one-proton separation energy ($S_{p}$) or two-proton separation energy ($S_{2p}$) for isotopes near the proton drip line. New facilities for radioactive ion beams make it possible to search for the location of the neutron drip line. Besides the  $\sim$3000 isotopes which have been found experimentally~\cite{DiscIsot}, most isotopes near the drip lines are only theoretically predicted to exist \cite{AME16}. These rare isotopes near the drip lines test the limits of both radioactive ion beam facilities and detector systems, since they have very low production probabilities in experiments. Thus it is always important to estimate the production of rare isotopes in experiments. Empirical parameterizations including {\sc epax3} \cite{epax3}, {\sc fracs} \cite{fracs} (which is an improved version of {\sc epax3}), and {\sc spacs} \cite{spacs}, etc., have been developed to estimate the cross sections of rare isotopes in projectile fragmentation and spallation reactions.
Besides the evaluation of binding energy and separation energy for rare isotopes, many theoretical methods have been developed to predict the binding energy of rare isotopes by extending the basic mass formula, for example, the macroscopic-microscopic approach \cite{MMSob}, the Weizs\"{a}cker-Skyrme formula \cite{WS10PRC1,WS10PRC2,WS11PRC,WangN14PLB}, and the improved J\"{a}necke mass formula \cite{Zhao14Mass}. Some empirical formulas have also been found via the correlation between the yield of fragments and their free energy or binding energy \cite{Tsang07BET,MFM1,Song17JPG}. In this paper, a method is proposed to determine the binding energies $S_n$ and $S_{2n}$ from isotopic yield.

%The cross section for fragments in the 140$A$ MeV $^{40, 48}$Ca + $^{9}$Be/$^{181}$Ta and $^{58, 64}$Ni + $^{9}$Be/$^{181}$Ta reactions have been measured by M. Mocko \textit{et al} at the National Superconducting Cyclotron Laboratory (NSCL) in Michigan State University \cite{Mocko06}.

\section{Formulism}
%\label{method}
Tsang \textit{et al} have proposed a method to determine $B$ for the very neutron-rich copper isotopes, for which the isotopic distribution depends exponentially on the average binding energy per nucleon \cite{Tsang07BET,MockoExpDep07},
\begin{equation}\label{BvsSigma}
\sigma = C \mbox{exp}[(B' - 8)/\tau],
\end{equation}
where $B' = (B - \varepsilon_{p})/A$, with $\varepsilon_{p}$ being the pairing energy. $C$ and $\tau$ are free parameters.
$\varepsilon_{p}$ is introduced to minimize the odd-even staggering in the isotopic distribution, which is:
\begin{equation}\label{Ap1}
\varepsilon_{p} = 0.5[(-1)^N + (-1)^Z]\varepsilon\cdot A^{-3/4}.
\end{equation}
$\varepsilon =$ 30 MeV is adopted, as done in Ref.~\cite{Tsang07BET}.
For an isotope with charge and mass numbers $(Z, A)$, taking the logarithm of Eq. (\ref{BvsSigma}) and multiplying by $A$, one has:
\begin{eqnarray}\label{ASigma}
&A\mbox{ln}\sigma_{(Z, A)}= A\mbox{ln}C \hspace{5.5cm}\nonumber\\
& + [B_{(Z, A)} - \varepsilon_{p(Z, A)} - 8A]/\tau,
\end{eqnarray}
where $(Z,A)$ is used as an index to indicate the isotope. For the isotope $(Z, A-1)$, i.e., one neutron removed from the isotope $(Z, A)$,
\begin{eqnarray}\label{A1Sigma}
&(A-1)\mbox{ln}\sigma_{(Z, A-1)}= (A-1)\mbox{ln}C \hspace{3.5cm}\nonumber\\
& + [B_{(Z, A-1)} - \varepsilon_{p(Z, A-1)} - 8(A-1)]/\tau,
\end{eqnarray}
\begin{eqnarray}\label{A2Sigma}
&(A-2)\mbox{ln}\sigma_{(Z, A-2)} = (A-2)\mbox{ln}C \hspace{3.5cm}\nonumber\\
& + [B_{(Z, A-2)} - \varepsilon_{p(Z, A-2)} - 8(A-2)]/\tau.
\end{eqnarray}
Combining Eqs. (\ref{ASigma}) and (\ref{A1Sigma}), for the isotope $(Z, A-1)$, i.e., one neutron removed from the isotope $(Z, A)$,
\begin{eqnarray}\label{SnSigma}
&A\mbox{ln}\sigma_{(Z, A)}-(A-1)\mbox{ln}\sigma_{(Z, A-1)} \hspace{3.8cm}\nonumber\\
&=\mbox{ln}C+[B_{(Z,A)}-B_{(Z,A-1)}-\varepsilon_{p(Z, A)}+\varepsilon_{p(Z, A-1)}\hspace{1cm}\nonumber\\
&-8]/\tau  \hspace{6.8cm}
\end{eqnarray}
Similarly, combining Eqs. (\ref{ASigma}) and (\ref{A2Sigma}), for the isotope $(Z, A-2)$, i.e., two neutrons removed from the isotope $(Z, A)$,
\begin{eqnarray}\label{S2nSigma}
&A\mbox{ln}\sigma_{(Z, A)}-(A-2)\mbox{ln}\sigma_{(Z, A-2)} \hspace{4cm}\nonumber\\
&=2\mbox{ln}C+[B_{(Z,A)}-B_{(Z,A-2)}-\varepsilon_{p(Z, A)}+\varepsilon_{p(Z, A-2)}\hspace{1.2cm}\nonumber\\
&-16]/\tau  \hspace{6.8cm}
\end{eqnarray}

The definitions of one-neutron separation energy and two-neutron separation energy for an isotope are:
\begin{equation}\label{SnDef}
S_n^{(Z, A)}=B_{(Z, A)} - B_{(Z, A-1)}
\end{equation}
and
\begin{equation}\label{S2nDef}
S_{2n}^{(Z, A)}=B_{(Z, A)} - B_{(Z, A-2)},
\end{equation}
respectively. The definitions of $S_n$ and $S_{2n}$ in relation to binding energy are the same as those in relation to atomic mass \cite{AME16}. Inserting Eq. (\ref{SnDef}) into (\ref{SnSigma}), with the definition of $\sigma_{(Z,A)}^{(-n)}\equiv A\mbox{ln}\sigma_{(Z, A)}-(A-1)\mbox{ln}\sigma_{(Z, A-1)}$, one has
\begin{equation}\label{SntoSigma}
\sigma_{(Z,A)}^{(-n)}=\mbox{ln}C+[S_n^{(Z, A)}-\varepsilon_{p(Z, A)}+\varepsilon_{p(Z, A-1)}-8]/\tau,
\end{equation}
Similarly, inserting Eq. (\ref{S2nDef}) into (\ref{S2nSigma}), with the definition of $\sigma_{(Z,A)}^{(-2n)}\equiv A\mbox{ln}\sigma_{(Z, A)}-(A-2)\mbox{ln}\sigma_{(Z, A-2)}$, one has
\begin{equation}\label{S2ntoSigma}
\sigma_{(Z,A)}^{(-2n)} = 2\mbox{ln}C + [S_{2n}^{(Z, A)} - \varepsilon_{p(Z, A)} + \varepsilon_{p(Z, A-2)} - 16]/\tau,
\end{equation}
For simplification, $S_{n}' \equiv S_{n}^{(Z, A)} - \varepsilon_{p(Z, A)} + \varepsilon_{p(Z, A - 1)}$ and $S_{2n}' \equiv S_{2n}^{(Z, A)} - \varepsilon_{p(Z, A)} + \varepsilon_{p(Z, A - 2)}$ are defined. By Eqs. (\ref{SnSigma}), (\ref{S2nSigma}), (\ref{SntoSigma}), and (\ref{S2ntoSigma}), $\sigma$, $B$, or $S_{n}$ ($S_{2n}$) can be predicted from each other based on known parameters. The parameters $C$ and $\tau$ in this work are determined using the least squares method. The cross sections, as well as the uncertainties, predicted by experimental $S_n$ and $S_{2n}$, adopt the prediction technique; $S_{n}$ or $S_{2n}$ and the uncertainties are determined from the cross section using the reverse prediction technique based on the least squares method \cite{lsm}.

\begin{figure}[htbp]
%\centering
\includegraphics%[width=\columnwidth]
[width=8.6cm]{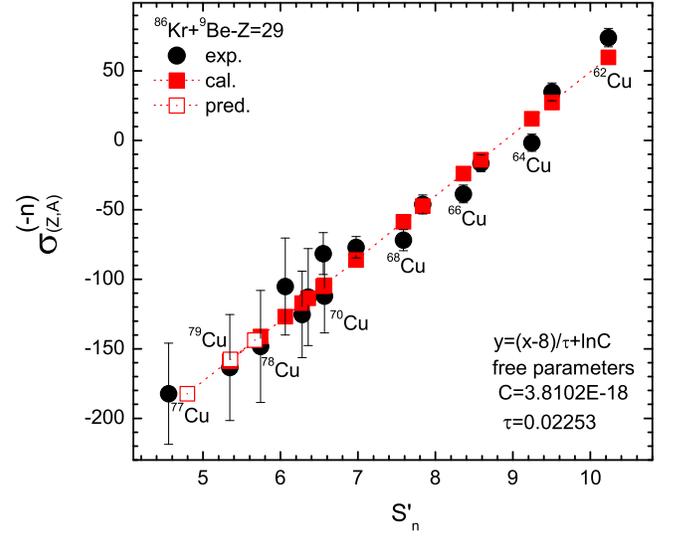}
\caption{\label{SnPlot} (Color online) The correlation between $\sigma_{(Z,A)}^{(-n)}$ and $S_{n}'$ for the measured isotopes in the 64$A$ MeV $^{86}$Kr + $^9$Be reaction. The circles denote $S_{n}^{'}$ and are calculated from the measured $S_{n}$ in AME16 \cite{AME16}. The line denotes the fitting to the $\sigma_{(Z,A)}^{(-n)}$ and $S_{n}'$ correlation for the measured results according to Eq. (\ref{SntoSigma}). The full squares denote the calculated $\sigma_{(Z,A)}^{(-n)}$ using the fitting function and the experimental $S_n$. The open squares denote the $S_{n}^{'}$ predicted from the fitting function by $\sigma_{(Z,A)}^{(-n)}$ with no measured $S_{n}$ in AME16.
}
\end{figure}

\begin{figure}[htbp]
%\centering
\includegraphics%[width=\columnwidth]
[width=8.6cm]{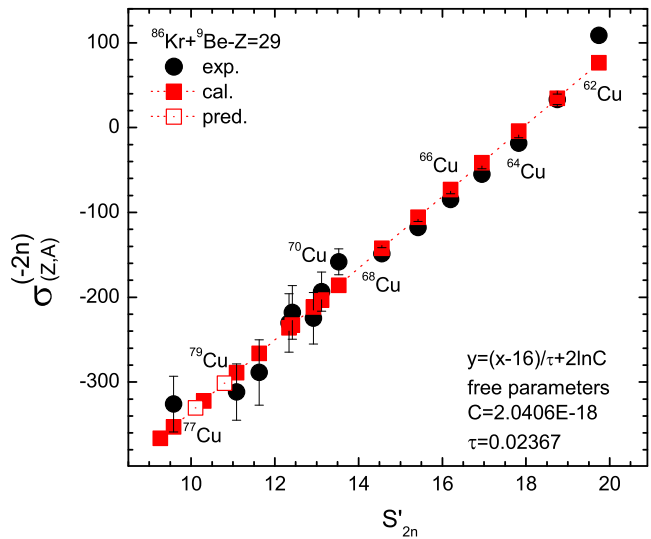}
\caption{\label{S2nPlot} (Color online) The correlation between $\sigma_{(Z,A)}^{(-2n)}$ and $S_{2n}'$ for the measured fragments in the 64$A$ MeV $^{86}$Kr + $^9$Be reaction. The circles denote $S_{2n}^{'}$ and are calculated from the measured $S_{2n}$ in AME16. The line denotes the fitting to the $\sigma_{(Z,A)}^{(-2n)}$ and $S_{2n}'$ correlation for the measured results according to Eq. (\ref{S2ntoSigma}). The full squares denote the calculated $\sigma_{(Z,A)}^{(-2n)}$ using the fitting function and the experimental $S_{2n}$. The open squares denote the $S_{2n}^{'}$ predicted from the fitting function by $\sigma_{(Z,A)}^{(-2n)}$.
}
\end{figure}

The cross sections for the neutron-rich copper isotopes in the 64 MeV/u $^{86}$Kr + $^9$Be reaction \cite{Tsang07BET}, which were measured at RIKEN by Tsang \textit{et al}, are adopted to perform the analysis. The values for $C$ and $\tau$ have been determined to be $C =$ 2.17 $\times$ 10$^{-15}$ mb and $\tau =$ 0.0213 MeV \cite{Tsang07BET}. The results of $S_{n}$ and $S_{2n}$ for the copper isotopes in the new version of the Atomic Mass Evaluation (AME16) \cite{AME16} are adopted. The correlation between $\sigma_{(Z,A)}^{(-n)}$ and $S_{n}'$ for the neutron rich copper isotopes is plotted in Fig. \ref{SnPlot}, from which a quite a good linear correlation can be found. The fitting to the $\sigma_{(Z,A)}^{(-n)}$ and $S_{n}'$ correlation yields $C =$ 3.81 $\times$ 10$^{-18}$ mb and $\tau =$ 0.0225 MeV. The fitted $\tau$ is similar to that in Ref. \cite{Tsang07BET}, while $C$ is much smaller than that in Ref. \cite{Tsang07BET}. The correlation between $\sigma_{(Z,A)}^{(-2n)}$ and $S_{2n}'$ for the copper isotopes is plotted in Fig. \ref{S2nPlot}, showing that it obeys the theoretical prediction much better than $\sigma_{(Z,A)}^{(-n)}$ and $S_{n}'$. Meanwhile, the odd-even staggering is less obvious compared to $\sigma_{(Z,A)}^{(-n)}$.

In Table \ref{SnS2nVrfy}, the results for $S_n$ and $S_{2n}$ of $^{65 - 76}$Cu isotopes obtained by Eqs. (\ref{SnDef}) to (\ref{S2ntoSigma}) are compared to the experimental data in AME16 \cite{AME16}. It can be seen that $S_n$ obtained by Eq. (\ref{SnDef}) is very similar to the AME16 data. $S_n$ obtained by Eq. (\ref{SntoSigma}) is very close to the AME16 data except for some isotopes which deviate  from the fitting line. The largest difference between $S_n$ from Eq. (\ref{SntoSigma}) and AME16 data is no more than $\pm$0.5 MeV, and most are within $\pm$0.3 MeV. A similar result is found for $S_{2n}$ determined by Eqs. (\ref{S2nDef}) and the AME16 data. The results suggest that $S_n$ and $S_{2n}$ determined by the combined isotopic cross sections are reasonable and have a high quality.

\section{Results and discussion}
%\label{RAD}

\begin{table}[thbp]
\caption{The results for $S_n$, $S_{2n}$ determined by Eqs. (\ref{SnDef}), (\ref{S2nDef}), (\ref{SntoSigma}) and (\ref{S2ntoSigma}) compared with AME16 \cite{AME16} data (in MeV) for $^{65-76}$Cu.}
\label{SnS2nVrfy}
\centering
\begin{tabular}{p{25pt}p{30pt}p{35pt}p{30pt}p{2pt}p{30pt}p{35pt}p{30pt}}
  \hline
  \hline
           &   &$S_n$   &      &  &      &$S_{2n}$    &    \\
                   \cline{2-4}               \cline{6-8}
  nuclei   & Eq. (\ref{SnDef}) &Eq. (\ref{SntoSigma})  &AME16 &  &Eq. (\ref{S2nDef}) &Eq. (\ref{S2ntoSigma})  &AME16\\
  \hline
  $^{65}$Cu &9.1904  &9.8583   &9.9104   &   &17.8266   &17.4970   &17.8265  \\
  $^{66}$Cu &7.0659  &6.7399   &7.0659   &   &16.9764   &16.6566   &16.9764  \\
  $^{67}$Cu &9.1325  &9.1617   &9.1326   &   &16.1985   &15.9262   &16.1985  \\
  $^{68}$Cu &6.3188  &6.0183   &6.3188   &   &15.4514   &15.1666   &15.4514  \\
  $^{69}$Cu &8.2406  &8.4376   &8.2405   &   &14.5594   &14.4073   &14.5593  \\
  $^{70}$Cu &5.3115  &5.8291   &5.3115   &   &13.5520   &14.2071   &13.552  \\
  $^{71}$Cu &7.8060  &7.6192   &7.8061   &   &13.1175   &13.3487   &13.1175  \\
  $^{72}$Cu &5.1432  &5.1495   &5.1432   &   &12.9493   &12.6334   &12.9493  \\
  $^{73}$Cu &7.2757  &7.7499   &7.2758   &   &12.4190   &12.7721   &12.4189  \\
  $^{74}$Cu &5.0899  &4.8923   &5.09     &   & 12.3656  &12.5006   &12.366  \\
  $^{75}$Cu &6.5363  &6.4106   &6.536    &   &11.6262   &11.0947   &11.627  \\
  $^{76}$Cu &4.5765  &4.3968   &4.576    &   &11.1128   &10.5730   &11.112  \\
  \hline\hline
\end{tabular}\\
\end{table}
For $^{77, 78, 79}$Cu, the values of $S_n$ in AME16 are estimated, as there are no experimental results. The values of $S_{n}$ determined for these isotopes are listed in Table \ref{SnS2nBoA}, and are a little larger than the evaluated results in AME16.
The results for $S_{2n}$ for $^{77, 79}$Cu are determined from $\sigma_{(Z,A)}^{(-2n)}$. $S_{2n}$ for $^{77}$Cu is close to the evaluated result in AME16, but for $^{79}$Cu it is 1.54 MeV larger than that in AME16. Combining the determined $S_n$ and $S_{2n}$, it is predicted that copper should have more neutron-rich isotopes than $^{79}$Cu.

\begin{table}[thbp]
\caption{The results for $S_n$, $S_{2n}$ and $B/A$ (in MeV) for the neutron-rich copper isotopes determined (Prd.) in this work, in Ref. \cite{Tsang07BET}, and evaluated in AME16 \cite{AME16}.}
\label{SnS2nBoA}
\centering
\begin{tabular}{p{20pt}p{22pt}p{22pt}p{2pt}p{22pt}p{22pt}p{2pt}p{22pt}p{22pt}p{22pt}p{22pt}}
  \hline
  \hline
    &    &$S_n$     & &   &$S_{2n}$ &    & &$B/A$ &&\\
  \cline{2-3} \cline{5-6}\cline{8-11}
            & Prd. &\cite{AME16}  & &Prd.  &\cite{AME16}      &   &Prd.$^*$    &Prd.$^{\dag}$ &\cite{Tsang07BET}&\cite{AME16}\\
  \hline
  $^{77}$Cu &5.961 &5.72 & &10.10 &10.3        &   &8.411       &8.406     &8.404    &8.408\\
  $^{78}$Cu &4.523 &3.95 & &10.24 &9.6$^{\S}$  &   &8.362       &8.358     &8.354    &8.351$^{\S}$\\
  $^{79}$Cu &6.497 &5.31 & &10.80 &9.26        &   &8.328       &8.330     &8.327    &8.312\\
  \hline\hline
\end{tabular}\\
$^*$ and $^{\dag}$ denote $B/A$ is determined by $S_n$ and $S_{2n}$, respectively. $^{\S}$ denotes an experimental result in AME16.
\end{table}

The binding energy per nucleon ($B/A$) for these isotopes is also calculated from the values of $S_n$ and $S_{2n}$ determined in this work using Eqs. (\ref{SnDef}) and (\ref{S2nDef}). In Table \ref{SnS2nBoA} it can be seen that the values of $B/A$ determined from  $S_n$ and $S_{2n}$ are very close to the AME16 evaluation.

\section{Summary}
%\label{summary}
Equations (\ref{SnDef}) and (\ref{SntoSigma}) both provide methods to determine $S_n$, and Eqs. (\ref{S2nDef}) and (\ref{S2ntoSigma}) both provide the methods to determine $S_{2n}$ for neutron-rich isotopes. For the very neutron-rich isotopes, it is not easy to measure the binding energy due to the low probability in experiments, which makes it difficult to determine $S_n$ or $S_{2n}$ from binding energy using Eqs. (\ref{SnDef}) and  (\ref{S2nDef}). Using Eqs. (\ref{SntoSigma}) and (\ref{S2ntoSigma}), $S_n$ or $S_{2n}$ can be determined from the yields or cross sections of isotopes, which are usually measured in projectile fragmentation reactions with high quality. Moreover, the neutron separation energy is linked to the neutron skin of neutron-rich isotopes \cite{Snskin1}. The predicted $S_{n}$, $S_{2n}$ or $\sigma$ can also be used to guide the adjustment of theoretical calculations related to neutron-skin thickness \cite{nskinSig1,nskinSig2,nskinSig3,nskinSig4}.

The method proposed in this paper indicates that $S_n$ and $S_{2n}$ of different isotopes can be determined by the combined isotopic cross sections. Furthermore, this method can also be used to find the location of neutron-drip isotopes from known cross sections, and vice versa to predict the production cross section of an isotope. This can help to design experiments to search for the location of the neutron-drip line.

%\begin{acknowledgments}
\textit{This work is partially supported by the Program for Science and Technology Innovation Talents at Universities of Henan Province (13HASTIT046), the Natural and Science Foundation in Henan Province (grant No. 162300410179), and the Program for the Excellent Youth at Henan Normal University (grant No. 154100510007). Y-D Song thanks the support from the Creative Experimental Project of National Undergraduate Students (CEPNU 201510476017).}
%\end{acknowledgments}

\end{document}